\documentclass[aps,prl,superscriptaddress,twocolumn]{revtex4-1}
\usepackage{amssymb,amsmath}
\usepackage{graphicx}
\usepackage{braket}
\newcommand{\EP}{\mathcal{F}}

\begin{document}

\title{Efficiencies of Quantum Optical Detectors}
\author{Daniel Hogg}
\affiliation{Institute for Quantum Science and Technology, University of Calgary, Alberta T2N1N4, Canada}
\author{Dominic W. Berry}
\affiliation{Department of Physics and Astronomy, Macquarie University, Sydney, NSW 2109, Australia}
\author{A. I. Lvovsky}
\affiliation{Institute for Quantum Science and Technology, University of Calgary, Alberta T2N1N4, Canada}
\affiliation{Russian Quantum Centre, 100 Novaya St., Skolkovo,
Moscow 143025, Moscow, Russia}
\email{LVOV@ucalgary.ca}
\date{\today}

\begin{abstract}
We propose a definition for the efficiency that can be universally applied to all classes of quantum optical detectors. This definition is based on the maximum amount of optical loss that a physically plausible device can experience while still replicating the properties of a given detector. We prove that detector efficiency cannot be increased using linear optical processing. That is, given a set of detectors, as well as arbitrary linear optical elements and ancillary light sources, it is impossible to construct detection devices that would exhibit higher efficiencies than the initial set.
\end{abstract}


\maketitle

Optical detectors --- devices for converting optical signals into electric ones --- are paramount not only in physics, but also in many aspects of our everyday life. In quantum optics and its applications to information processing and communication, detectors are particularly diverse and subject to intense study \cite{tomo,hybrid,review1,review2} and development \cite{BullerCollins,Achilles03}. This is necessary to satisfy the demands associated with various methods of quantum-optical state measurement required for different quantum technology applications.

A primary performance benchmark of any optical detector is its quantum efficiency. In spite of its universality and intuitiveness,  this fundamental characteristic does not have a uniform definition applicable to all classes of detectors. For example, the quantum efficiency of a photodiode is defined as a ratio of the number of photoelectrons to the number of incident photons; that of a single-photon detector is the probability to generate a ``click" given a single input photon; for a balanced homodyne detector, the efficiency is obtained by means of a relatively complex calculation that includes the efficiencies of its photodiodes, the mode matching of the signal and the local oscillator \cite{LvovskyRaymer}, the electronic noise \cite{Appel07} and other parameters.

A further important outstanding problem is the construction of optical detectors with high efficiency. Efficient optical detection is a primary requirement in applications ranging from quantum information processing and communications \cite{Varnava08,Scherer11,key2} to fundamental tests such as  loophole-free locality violation \cite{Giustina13,Christensen13}. Although significant progress has been made in recent years in photon detector technology \cite{BullerCollins}, highly efficient optical detectors remain expensive and unavailable for certain wavelengths. It would therefore be useful to develop means of increasing detector efficiency by optical means. That is, construct an all-optical device involving lower-efficiency detectors that behaves as a higher-efficiency detector.

It is possible to increase photon detection efficiency using non-linear optics. Examples include  non-demolition detection \cite{Munro05} and the CNOT gate \cite{cnot}. For quadrature detection, the efficiency can be enhanced via optical squeezing to amplify the quadrature that is in phase with the local oscillator \cite{squeezing}. However, nonlinear optical processing is  typically lossy and requires sophisticated technology. In contrast, linear optical elements with very low loss are routinely manufactured. It would be therefore much more preferable to increase the efficiency of a set of detectors using only linear optical elements.

In this paper, we introduce a generalized definition of detector efficiency. Thereafter, we are able to address the principal question of this paper: Can the efficiency of a set of detectors be increased using linear optics? We show that the answer to this question is negative.

The efficiency of a photon detector can be defined in
terms of equivalent optical loss \cite{Renema:12,Calkins:13}. That is, for a given detector, we look for equivalent configurations consisting of another detector preceded by a loss channel (attenuator) of transmissivity $\eta$ [Fig.~1(a)]. The detector efficiency would be the minimum (or infimum) value of $\eta$ for all such equivalent representations that are theoretically allowed.

Mathematically, the generalized efficiency of a detector with a positive-operator valued measure (POVM) $\vec{\hat{\Sigma}}$ is  given by
\begin{equation}\label{geneff}
E(\vec{\hat{\Sigma}}) = \text{inf}\left\{\eta | \exists \vec{\hat{\Pi}},\EP_{\eta}(\vec{\hat{\Pi}}) = \vec{\hat{\Sigma}}\  \right\},
\end{equation}
where $\EP_{\eta}(\vec{\hat{\Pi}})$ represents  a detector with POVM $\vec{\hat{\Pi}}$ with an attenuator of transmissivity $\eta$ placed in front of it [Fig.~1(a)]. $\vec{\hat{\Pi}}$ must be a theoretically allowed POVM, i.e. a set of non-negative self-adjoint operators that sum to unity.

For a detector whose POVM $\vec{\hat{\Pi}}$ is known, the loss transformation $\EP_{\eta}(\vec{\hat{\Pi}})$ can be calculated as follows. For an input quantum state $\hat{\rho}$, the probability of obtaining a specific ($\ell$th) measurement outcome
is
\begin{equation}\label{trace1}
p_\ell(\hat{\rho}) = \text{Tr}[\hat{\rho}\EP_\eta(\hat{\Pi}_\ell)].
\end{equation}
On the other hand, the detection process shown in Fig.~1(a) is equivalent to that in Fig.~1(b), so we can write
\begin{equation}
p_\ell(\hat{\rho}) = \text{Tr}[\mathcal{E}_\eta(\hat{\rho})\hat{\Pi}_\ell], \label{trace2}
\end{equation}
where $\mathcal{E}_\eta(\hat{\rho})$ is the loss transformation of state $\hat\rho$, which in the Fock basis takes the form of the  the generalized Bernoulli transformation \cite{lee,bernoulli}. This can then be used to derive $\EP_{\eta}(\vec{\hat{\Pi}})$ in that basis as shown in Appendix A. For a diagonal POVM, relevant for phase-insensitive detection, the map takes the form

\begin{equation}
\label{eq:digres}
\bra{n}\EP_{\eta}(\hat{\Pi}_\ell)\ket{n} = \sum_{k=0}^n \binom{n}{k} (1-\eta)^{n-k} \eta^k \bra{k}\hat{\Pi}_\ell\ket{k}.
\end{equation}

\begin{figure}[t]
\includegraphics[width=0.95\columnwidth]{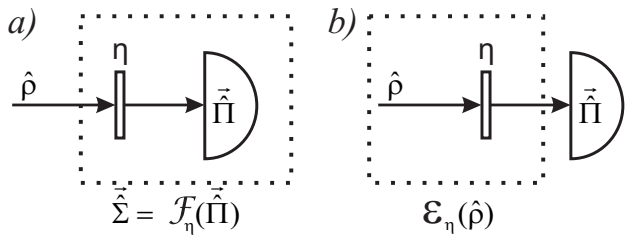}
\caption{Equivalent loss model of detector. (a) An inefficient detector with POVM $\vec{\hat{\Sigma}}$ and efficiency $\eta$ is equivalent to a detector with POVM $\vec{\hat{\Pi}}$ preceded by an attenuator with transmissivity $\eta$. (b) An equivalent model of quantum state measurement with an imperfect detector.}
\label{fig:eff}
\end{figure}

To illustrate the results above, let us consider a non-discriminating single-photon detector of efficiency $\eta$, defined as the probability for a single incident photon to generate a ``click". This detector is described by the well-known POVM \cite[p.~118]{kok}
\begin{subequations}\label{SPCM}
\begin{align}
\hat{\Pi}_{\text{off}}(\eta) &= \sum_{n=0}^\infty (1-\eta)^n \ket{n}\bra{n}, \\
\hat{\Pi}_{\text{on}}(\eta) &= \sum_{n=0}^\infty [1-(1-\eta)^n] \ket{n}\bra{n}.
\end{align}
\end{subequations}
If we place an attenuator with transmissivity $\eta'$ in front of that detector, transformation (\ref{eq:digres}) will lead to
\begin{subequations}
\begin{align}
\EP_{\eta'}(\hat{\Pi}_{\text{off}}(\eta)) &= \sum_{n=0}^\infty (1-\eta\eta')^n \ket{n}\bra{n}=\hat{\Pi}_{\text{off}}(\eta\eta'), \\
\EP_{\eta'}(\hat{\Pi}_{\text{on}}(\eta)) &= \sum_{n=0}^\infty [1-(1-\eta\eta')^n] \ket{n}\bra{n}=\hat{\Pi}_{\text{on}}(\eta\eta'),
\end{align}
\end{subequations}
i.e.~this setting is equivalent to a non-discriminating single-photon detector of efficiency $\eta\eta'$.

By the same token, a detector with POVM (\ref{SPCM}) is equivalent to a non-discriminating detector of efficiency $\eta/\eta'$ preceded by an attenuator with transmissivity $\eta'$. The POVM elements $\hat{\Pi}_{\text{off,on}}(\eta/\eta')$ are non-negative for $\eta'\ge\eta$, and negative (unphysical) for $\eta'<\eta$. This implies that the generalized efficiency (\ref{geneff}) of detector (\ref{SPCM}) equals $\eta$, so our new definition is consistent with the traditional one. One can use similar arguments to show this consistency for other types of detectors.

The equivalent-loss approach to quantum efficiencies has previously been applied to investigate the question of whether linear optical processing can increase the efficiency of single photon sources. Originally investigated in \cite{prl7,prl9}, an explicit definition of source efficiency was constructed, along with a proof that linear optical processing cannot increase it \cite{prl2010}. Later, this proof was extended to the case of multiple sources \cite{pra2011}. Below, we address a similar problem for detectors.

Suppose we are given a set of single-mode detectors with efficiencies  $\{\eta_i\}$, which we call ``physical".
One may use these detectors, an arbitrary number of linear optical elements, such as beam splitters and phase shifters, as well as any ancillary light sources, to construct a set of single-mode optical state measurement devices which we call ``virtual detectors" [Fig.~2(a)]. Different virtual detectors do not share any optical elements.
In particular, each virtual detector uses a different subset of the physical detectors.
We show that the efficiencies $\eta'_i$ of these virtual detectors are bounded by the efficiencies of the physical detectors:
\begin{equation}\label{statement}
 {\eta'_i}^\downarrow\le \eta_i^\downarrow,
\end{equation}
where the downward arrow denotes sorting in non-increasing order.

Consider first a single virtual detector. It can be represented by a scheme shown in Fig.~2(a). The mode $\hat a_1$ to be measured, as well as the ancillary modes  $\hat a_{2,\ldots,M}$, are processed by an interferometer. Such an interferometer can be represented in the Heisenberg picture as a unitary transformation $W$ of the input and output mode annihilation operators.
\begin{equation}\label{W}
\hat a_i = \sum_{j=1}^{M} W_{ij} \hat{b}_j.
\end{equation}
The output modes $\{\hat{b}_j\}$ may be incident onto the physical detectors or simply discarded. The latter case requires no special treatment because discarding a mode is equivalent to measuring it with a detector with efficiency zero.

\begin{figure}[t]
\includegraphics[width=\columnwidth]{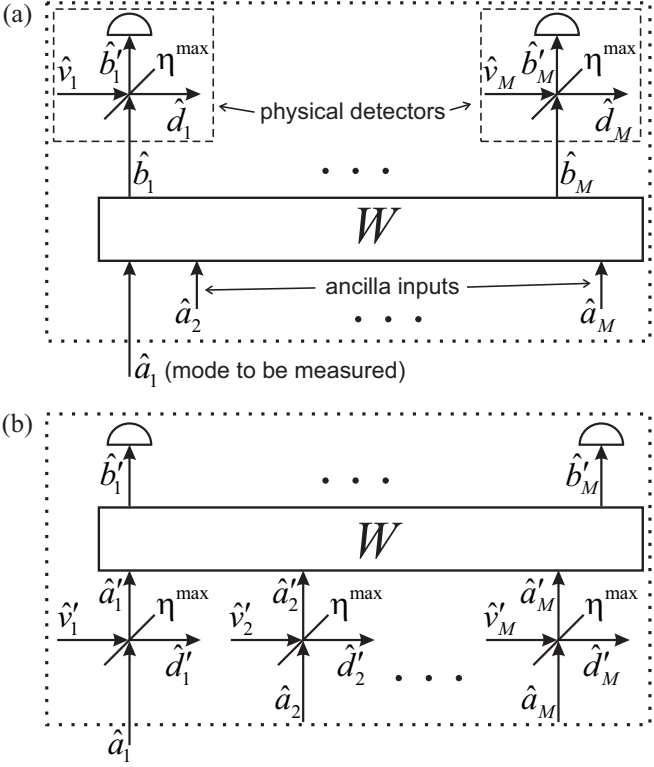}
\caption{A single-mode virtual detector. (a) A generalized model. The mode to be measured and the ancillary modes are processed by an interferometer and impinge onto the physical detectors, whose equivalent models are shown inside the dotted rectangles. (b) An equivalent model of the virtual detector. The input mode is subjected to immediate attenuation, which implies that the detector efficiency cannot exceed $\eta^{\max}$.}
\label{setup}
\end{figure}

Let $\eta^{\max}$ be the highest efficiency of all physical detectors used in a given virtual detector. According to our definition of detector efficiency, the physical detectors are equivalent to higher-efficiency detectors preceded by attenuators of transmissivity $\eta^{\max}$.
As our definition of efficiency uses an infimum, the transmissivity may need to be taken as $\eta^{\max}+\epsilon$ for arbitrarily small (but positive) $\epsilon$.
It is trivial to include this $\epsilon$, and it does not qualitatively change the proof.
We therefore omit it in the following discussion.
As demonstrated in Ref.~\cite{prl2010}, the attenuators can be commuted to precede the interferometer. We now present a simpler argument to that effect.

Using the beam splitter model of loss \cite{Leonhardt}, we decompose the modes incident on them as
\begin{eqnarray}\label{outBSb}
\hat{b}_j &=& \sqrt{\eta^{\max}} \hat{b}'_j + \sqrt{1-\eta^{\max}} \hat{d}_j,\\
\hat{v}_j &=& \sqrt{1-\eta^{\max}} \hat{b}'_j - \sqrt{\eta^{\max}} \hat{d}_j,\label{outBSv}
\end{eqnarray}
where modes  $\hat{d}_j$  are discarded. Putting Eqs.~(\ref{W}) and (\ref{outBSb}) together, we write for the incoming mode
\begin{equation}\label{afinal}
\hat{a}_i = \sqrt{\eta^{\max}} \hat a'_i + \sqrt{1-\eta^{\max}} \hat{d}'_i
\end{equation}
with
\begin{equation}\label{dprimeprime}\hat{a}'_i := \sum_{j=1}^N W_{ij}  \hat{b}'_j \textrm{~ and ~}\hat{d}'_i := \sum_{j=1}^N W_{ij}  \hat{d}_j.
\end{equation}
At the same time, Eq.~(\ref{outBSv}) can be rewritten as
\begin{eqnarray}\label{vW}
\hat{v}'_i :=\sum_{j=1}^N W_{ij}  \hat{v}_j =\sqrt{1-\eta^{\max}}\hat a'_i - \sqrt{\eta^{\max}} \hat{d}'_j.
\end{eqnarray}

These results imply that the setup of Fig.~2(a) is equivalent to that of Fig.~2(b). Indeed, Eqs.~(\ref{afinal}) and (\ref{vW}) signify a beam splitter transformation. Additionally, modes $\hat{v}'_i$ can be obtained from vacuum modes $\hat{v}_j$ by means of a linear optical transformation, so they must also be in the vacuum state. Modes $\hat{d}'_i$ are related in a similar fashion to modes $\hat{d}_j$, and so can be treated as discarded.

Our virtual detector is hence equivalent to a setup in which an attenuator with transmissivity  $\eta^{\max}$ is placed in front of input mode $\hat a_1$. This implies that $\eta^{\max}$ is an upper bound for the efficiency of this detector.

\begin{figure}[t]
\includegraphics[width=0.894\columnwidth]{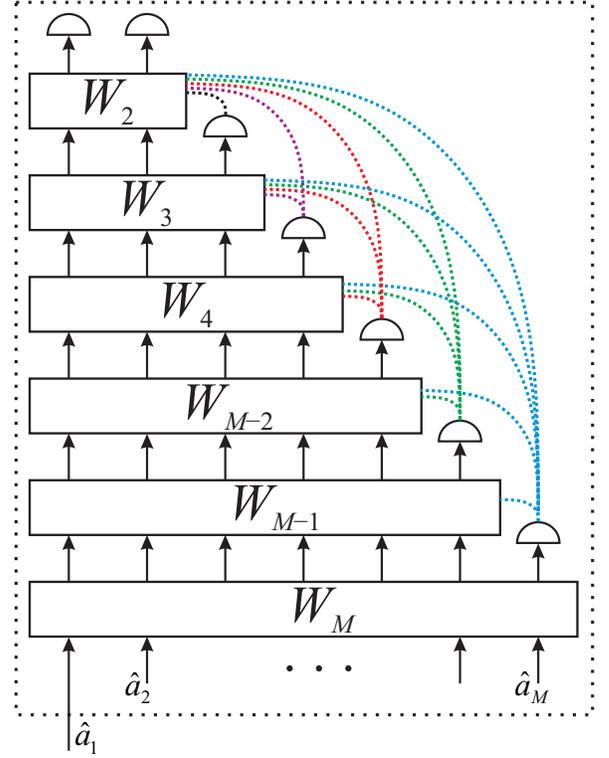}
\caption{(Color online) A single-mode virtual detector with adaptive measurements for $M=7$.
The virtual detector is again shown inside the dotted rectangle.
The dotted lines from the detectors to the interferometers indicate that the interferometers are controlled based on the results of the detections.}
\label{adaptive}
\end{figure}

This result is readily extended to virtual detectors with adaptive measurements;
that is, detectors in which the configuration of interferometer $W$ can be modified dependent on the results of the measurements by a subset of the physical detectors.
To see this, we model the adaptive virtual detector as shown in Fig.~\ref{adaptive}.
We would have the interferometer $W_M$ on modes $1,\ldots,M$, followed by interferometer $W_{M-1}$ on modes $1,\ldots,M-1$, and so forth up to interferometer $W_2$ on modes $1$ and $2$.
Each interferometer $W_k$ on modes $1,\ldots,k$ would depend on the results of measurements of modes $k+1,\ldots,M$.
This interferometer would then be followed by a measurement on mode $k$.

As discussed above, each physical detector can be modeled as having a beam splitter with transmissivity $\eta_{\max}$ before it.
In the same way as above, the beam splitters after interferometer $W_2$ can be commuted to before $W_2$, so that there are three equal reflectivity beam splitters after interferometer $W_3$.
These can then be commuted through $W_3$, and so forth, until we have beam splitters with transmissivity $\eta_{\max}$ before the first interferometer $W_M$.
Although the interferometers $W_2$ to $W_{M-1}$ can depend on measurement results, the loss commutes independently of the interferometers.
This means that the above argument holds, and the efficiency for this virtual detector cannot exceed $\eta_{\max}$.

We now proceed to proving Eq.~(\ref{statement}) for multiple virtual detectors.
We have so far shown that the efficiency $\eta'_i$ of the $i$th virtual detector cannot exceed the efficiency $\eta^{\max}_i$ of the best physical detector used in its construction; that is, $\eta'_i\le \eta^{\max}_i$.
Let $\sigma(i)$ and $\tau(i)$ be the permutations that define the sorting of sequences $\eta^{\max}_i$ and $\eta'_i$ in non-increasing order: $\eta^{\max\downarrow}_{\sigma(i)}=\eta_i^{\max}$ and ${\eta'_i}^\downarrow=\eta'_{\tau(i)}$, respectively.
Then we obtain
\begin{equation}
\label{easy}
{\eta'_i}^\downarrow = \eta'_{\tau(i)} \le \eta_{\tau(i)}^{\max} = \eta_{\sigma(\tau(i))}^{\max\downarrow}
\end{equation}
Now consider the case that Eq.~(\ref{statement}) were violated; that is
${\eta'_i}^\downarrow > \eta_i^\downarrow$.
That would imply that ${\eta'_k}^\downarrow > \eta_i^\downarrow$ for all $k\le i$,
which would in turn imply that $\eta_{\sigma(\tau(k))}^{\max\downarrow} > \eta_i^\downarrow$ for all $k\le i$.
But, that would imply that there are $i$ values of $\eta_k^\downarrow$ that are larger than $\eta_i^\downarrow$, which is a contradiction, because $\eta_i^\downarrow$ is sorted in non-ascending order.
This contradiction implies that Eq.~(\ref{statement}) must hold.


Linear optics are cheap and easy to manufacture. In addition, their properties are well-understood, and linear processes in general have high efficiency. It would be extremely fortunate if we could somehow use these processes to increase the efficiency of sources or detectors, but this is not the case. Nonlinear optics appears to be the only alternative.


This work has been supported by NSERC. We thank B. C. Sanders and A. S. Prasad for helpful discussions.
DWB is funded by an Australian Research Council Future Fellowship (FT100100761).
\bibliography{DetectorEfficiency19}

\appendix
\section{Appendix A: POVM loss transformation}
\label{sec:app}
Here we determine an expression for the map $\EP_\eta$ on the detector POVM under loss.
For full generality, we derive the transformation for the case where the POVM is not diagonal, so elements can be written in the Fock basis as
\begin{equation}
\hat{\Pi}_\ell = \sum_{m,n=0}^\infty (\hat{\Pi}_\ell)_{mn} \ket{m}\bra{n}.
\end{equation}
Using conditions \eqref{trace1} and \eqref{trace2} on $\EP_\eta$, taking the trace over both sides in the Fock basis and inserting identity matrices yields
\begin{align}
&\sum_{m=0}^\infty \sum_{n=0}^\infty \bra{m} \hat{\rho} \ket{n} \bra{n} \EP_\eta(\hat{\Pi}_\ell) \ket{m} \nonumber \\
&=\sum_{m=0}^\infty \sum_{n=0}^\infty \bra{m} \mathcal{E}_\eta(\hat{\rho}) \ket{n} \bra{n} \hat{\Pi}_\ell \ket{m} .
\end{align}
Substituting the expression for $\mathcal{E}_\eta(\hat{\rho})$ from the generalized Bernoulli transformation \cite{lee,bernoulli} then gives
\begin{widetext}
\begin{align}
&\sum_{m=0}^\infty \sum_{n=0}^\infty \bra{m} \hat{\rho} \ket{n} \bra{n} \EP_\eta(\hat{\Pi}_\ell) \ket{m} \nonumber \\
&= \sum_{m=0}^\infty \sum_{n=0}^\infty \sum_{k=0}^\infty \bra{m+k} \hat{\rho} \ket{n+k} \sqrt{\binom{m+k}{k}\binom{n+k}{k}}(1-\eta)^k \eta^{\frac{1}{2}(m+n)} \bra{n} \hat{\Pi}_\ell \ket{m} \nonumber \\
&=\sum_{m=0}^\infty \sum_{n=0}^\infty \sum_{k=0}^{\min(m,n)}\bra{m} \hat{\rho} \ket{n} \sqrt{\binom{m}{k}\binom{n}{k}}(1-\eta)^k \eta^{\frac{1}{2}(m+n)-k} \bra{n-k} \hat{\Pi}_\ell \ket{m-k}
\end{align}
Since the elements $\bra{m} \hat{\rho} \ket{n}$ are arbitrary, $\EP_\eta$ yields a transformation described by
\begin{equation}
\bra{n} \EP_\eta(\hat{\Pi}_\ell) \ket{m} = \sum_{k=0}^{\min(m,n)}  \sqrt{\binom{m}{k}\binom{n}{k}}(1-\eta)^k \eta^{\frac{1}{2}(m+n)-k} \bra{n-k} \hat{\Pi}_\ell \ket{m-k}. \label{general}
\end{equation}
\end{widetext}

\end{document}